\documentclass[twocolumn,aps,showpacs,prl,epsf,graphics,psfig]{revtex4}
\usepackage{graphicx}

\begin{document}
\title{Protection Strategy of Sensitive Body Organs 
in Radiation Therapy}

\author{Ramin M. Abolfath}
\author{Lech Papiez}

\affiliation{
Department of Radiation Oncology, University of Texas,
Southwestern Medical Center, Dallas, TX 75390
}

\date{\today}

%%%%%%%%%%%%%%%%%%%%%%%%%%%%%%%%%%%%%%%%%%%%%%%%%%%%%%%%%%%%%%%%%%%%%%%%%%
\begin{abstract}
In this paper, we investigate protection strategies of sensitive body anatomy
against the irradiation to the cancerous moving tumors in
intensity modulated radiation therapy.
Inspired by optimization techniques 
developed in statistical physics and dynamical systems,  
we deploy a method based on variational principles
and formulate an efficient genetic algorithm which enable us to
search for global minima in a complex landscape
of irradiation dose delivered to the radiosensitive organs at risk.
We take advantage of the internal motion of body anatomy during radiation 
therapy to reduce the unintentional delivery of the radiation to sensitive
organs.
We show that the accurate optimization of the control parameters,
compare to the conventional IMRT and widely used delivery 
based on static anatomy assumption, 
leads to a significant reduction of the dose delivered to the organs at 
risk.
\end{abstract}
\maketitle
%%%%%%%%%%%%%%%%%%%%%%%%%%%%%%%%%%%%%%%%%%%%%%%%%%%%%%%%%%%%%%%%%%%%%%%%%%
{\em Introduction:}
Optimality rules in nature play crucial role in characterizing
the dynamical pathway of complex systems such as
enzymatic reactions \cite{Lomholt}, 
protein folding \cite{Pande2000:RMP},
DNA repair mechanisms and carcinogenesis
\cite{Shih2008:PRL}
as the trajectories of the constituent atoms and molecules 
of these systems follow the 
principle of least action \cite{GoldsteinBook}.
Optimization techniques developed in recent years in statistical 
and computational physics have shown to be also
useful tools in facilitating treatment delivery methods,
in particular, in intensity modulated radiation therapy (IMRT) 
\cite{Khan:book,Webb:book,Bortfeld2006:MB,Laub2000:PMB,Hou2001:PRL,Cotrutz2003:PMB}.
IMRT is an inverse planning approach 
and delivery that aims at shaping dose distributions better to anatomical 
features allowing to increase exposure of tumors and decrease the 
toxicity of the treatments \cite{Khan:book,Webb:book,Bortfeld2006:MB,Laub2000:PMB,Hou2001:PRL}.
Engineering treatment delivery methods 
and the control of the leaf-velocity 
\cite{Keall2001:PMB,Papiez04:MP,Papiez05:MP}
and intensity-dose-rate \cite{Papiez08:MP}
for moving and deforming body anatomy \cite{Rangaraj05:MP} 
during hypo fractionation and IMRT 
have opened up a broad and relatively unexplored frontier for 
research in computational physics with application to radiation oncology.

The unavoidable stochastic nature of body anatomy motion in living systems
is one of the challenging problems in  
radiation therapy and needs to be integrated as a part of the
state of the art treatment planning systems.
Although IMRT is very important technological achievement, however, 
it faces many challanges 
when it is planned for the delivery to the body anatomies that 
experience motion,
particularly for treatments of thoracic and abdominal sites.
In the presence of motion the dose planned and dose delivered do not match,
and thus the clinical results can be worse than pre-IMRT treatments. 
% \cite{remark2}.
To prevent the danger of creating secondary cancer in normal tissues 
due to this problem a proactive approach to utilize 
the organ motions has to be taken into account to lower the dose to sensitive 
structures without compromising the dose delivered to the target
\cite{Papiez07:MP,Papiez08:MP}. 
Diverse criteria for quality of 
treatment can be utilized, some increasing the efficiency of treatment 
and minimizing 
overall radiation exposure \cite{Papiez04:MP,Papiez05:MP,Rangaraj05:MP} 
while other decreasing cumulative 
dose distribution over organs at risk (OAR) to minimize treatment toxicity 
\cite{Papiez07:MP,Papiez08:MP}. 
The recently proposed algorithms aim to change the paradigm of  
IMRT to moving body anatomy.
Taking into account the fact that multiple delivery strategies are 
possible to moving body anatomy, 
the goal is to use the additional degrees of freedom in the dynamical 
system of body anatomy and 
multi-leaf collimator
(MLC) systems, resulting from motions of body organs, 
to the advantage of treatment. 
Thus our goal is not to achieve dose distributions in moving body anatomy 
as good as it is possible for static anatomy but be able to deliver 
dose distributions to moving body that are advantageous to any dose 
distribution possible for static anatomy.
The solution strategies can lead to complex optimization processes, 
many supporting diverse landscapes of multiple local minima. 
Among few possible procedures leading to achieving global solutions, 
or satisfactory local solutions, we explore explicitly dynamical pathways 
that breed offspring control parameters that efficiently converge to 
advantageous solutions. 
These so called genetic operations \cite{KozaBook}, 
such as reproduction, suppression, mutation, and the mating crossover 
lead to efficient solutions of our multi-parameter dynamic treatment 
optimization problems.

We present a strategy based on controlling the velocity of multi-leafs 
and intensity dose rate of beams in dynamic multi-leaf collimator (DMLC) 
deliveries that allow to redistribute beneficially the 
irradiation to the target and to organs at risk over subsequent phases 
of the body motion.
To simplify our computations, we assume in this study that motion of 
the body anatomy is known before delivery, though this assumption is 
not compulsory for our approach \cite{Papiez07:MP,Papiez08:MP}.
In particular by using the velocity of the leaves and the intensity 
dose rate as the 
control parameters in optimization process we propose a strategy that 
the dose delivered to the organ at risk can be reduced significantly 
relative to the delivery based on static body anatomy assumption.
	
%%%%%%%%%%%%%%%%%%%%%%%%%%%%%%%%%%%%%%%%%%%%%%%%%%%%%%%%%%%%%%%%%%%%%%%%%%%%
%%%%%%%%%%%%%%%%%%%%%%%%%%%%%%%%%%%%%%%%%%%%%%%%%%%%%%%%%%%%%%%%%%%%%%%%%%%%
%%%%%%%%%%%%%%%%%%%%%%%%%%%%%%%%%%%%%%%%%%%%%%%%%%%%%%%%%%%%%%%%%%%%%%%%%%%%
%%%%%%%%%%%%%%%%%%%%%%%%%%%%%%%%%%%%%%%%%%%%%%%%%%%%%%%%%%%%%%%%%%%%%%%%%%%%

{\em Method:}
In the following we present a method based on optimization algorithm 
that mixes minimization of the time of delivery 
(and thus minimization of the contaminating irradiation) and minimization 
of dose delivered to OAR for exposure of moving body anatomy.
We use velocity of the leaves and the intensity beam-dose rate 
(or intensity rate) $\alpha$ as control parameters in optimization.
We consider the motion of the target to be aligned with the movements of 
MLC leaves in one-dimension.

{\em Optimizing time of delivery:}
Refs. \cite{Papiez04:MP,Papiez05:MP,Rangaraj05:MP} investigated
the mathematical model for the time-optimal control of a single leaf pair
(or a set of multiple independent leaf pairs) 
for the case of moving rigid target.
They explored deliveries for time-optimal DMLC strategies subjected 
to constant intensity rate.
To preserve the integrity of the delivered intensity in these solutions,  
the following leaf velocity ($v_F$), the leading leaf velocity ($v_L$), and
the target velocity ($v_T$) have to satisfy the relation given below
\begin{eqnarray}
% v_L(x,m_L)=\frac{v_F(x,m_F) - v_T(m_F)}{1-I'(x)[v_F(x,m_F) - v_T(m_F)]} + v_T(m_L),
% \nonumber\\
v_F(x,m_F)=\frac{v_L(x,m_L) - v_T(m_L)}{1+I'(x)[v_L(x,m_L) - v_T(m_L)]} + v_T(m_F).
\label{eq1}
\end{eqnarray}
In Eq. (\ref{eq1}),
$m_L$ and $m_F$ denote the time measured in cumulative monitor unit (MU)
at which the leading and following leaves pass over a point
on the target denoted by $x$ (defined in the target frame of reference),
$I'(x)\equiv dI/dx$, and $m_F - m_L = I(x)$.
Here $I(x)$ is the planned intensity, intendend to be delivered to the target. 
The velocities are expressed in cm/MU, i.e., 
$v_\lambda \equiv dx_\lambda/dm$ 
where $\lambda \in \{T,L,F\}$ and
$x_{L(F)}$ is the position of the edge of the leading (following) leaf in
the laboratory frame of reference.
$x_T$ is the position of the target.
Eq. (\ref{eq1}) is invariant under variation of $\alpha$.
Because of the mechanical design of MLC,
leaves can not move faster than a maximum accessible velocity
$v_{\rm max}$.
This condition enforces a lower-bound $m_{\rm min}$ for
the final moment of delivery $m_f$ such that
$m_f \geq m_{\rm min}$.
$m_{\rm min}$ decreases with increasing $v_{\rm max}$.
To achieve a unique solution corresponding to 
$m_f=m_{\rm min}$ with constant $\alpha$,  
one of the leaves moves with the maximum velocity $v_{max}$, 
%over a given point in the target frame of reference 
while the other 
moves with appropriately matched velocity \cite{Papiez04:MP}.
It follows
%  
% \begin{eqnarray}
% v_F=
% \left\{\begin{array}{c}
% v_{\rm max}~~~~~~~~~~~~~~~~~~~~~~~~~~~~~~~~~~~~~~ {\rm for~ case~ 2}\\
% \frac{v_{\rm max} - v_T(m_L)}{1+I'(x)[v_{\rm max} - v_T(m_L)]} + v_T(m_F)
% ~~~~~{\rm for~ case~ 1}
% \end{array}\right.
% \label{eq2}
% \end{eqnarray}
% 
% \begin{eqnarray}
% v_L=
% \left\{\begin{array}{c}
% v_{\rm max}~~~~~~~~~~~~~~~~~~~~~~~~~~~~~~~~~~~~~~ {\rm for~ case~ 1}\\
% \frac{v_{\rm max} - v_T(m_F)}{1-I'(x)[v_{\rm max} - v_T(m_F)]} + v_T(m_L)
% ~~~~~{\rm for~ case~ 2}
% \end{array}\right.
% \label{eq3}
% \end{eqnarray}
% Here $v_{\rm max}$ is expressed in MU.
% Trajectories of the leaves can then be easily calculated by integrating
% the velocities
\begin{eqnarray}
\frac{dx_F}{dm_F}=
\left\{\begin{array}{c}
v_{\rm max}~~~~~~~~~~~~~~~~~~~~~~~~~~~~~~~~~~~~~~ {\rm for~ case~ 2}\\
\frac{v_{\rm max} - v_T(m_L)}{1+I'(x)[v_{\rm max} - v_T(m_L)]} + v_T(m_F)
~~~~~{\rm for~ case~ 1}
\end{array}\right.
\label{eq4}
\end{eqnarray}
\begin{eqnarray}
\frac{dx_L}{dm_L}=
\left\{\begin{array}{c}
v_{\rm max}~~~~~~~~~~~~~~~~~~~~~~~~~~~~~~~~~~~~~~ {\rm for~ case~ 1}\\
\frac{v_{\rm max} - v_T(m_F)}{1-I'(x)[v_{\rm max} - v_T(m_F)]} + v_T(m_L)
~~~~~{\rm for~ case~ 2}
\end{array}\right.
\label{eq5}
\end{eqnarray}
% Note that $x_F$, and $x_L$ are the coordinates in laboratory frame of reference.
Here $v_{\rm max}$ is expressed in MU.
Trajectories of the leaves can then be easily calculated by integrating
the velocities.
Conditions defined as case-1 and case-2 denote the appropriate subsets 
of the Cartesian product of spatial variables spanning the whole range of 
target in the motion, and the interval of time (exceeding the time 
of delivery). 
We refer to Refs. \cite{Papiez04:MP,Papiez05:MP,Rangaraj05:MP} for details.
Applying the transformation $dm/dt=\alpha$ or equivalently 
$m=\int_0^t dt' \alpha(t')$ in Eq.(\ref{eq1}) leads to 
leaf velocities expressed in physical time $t$ \cite{Papiez08:MP}
% \begin{equation}
% \frac{v_L(x,t_L)}{\alpha(t_L)}=
% \frac{\frac{v_F(x,t_F)}{\alpha(t_F)} - \frac{v_T(t_F)}{\alpha(t_F)}}{1-I'(x)[\frac{v_F(x,t_F)}{\alpha(t_F)} - \frac{v_T(t_F)}{\alpha(t_F)}]} + \frac{v_T(t_L)}{\alpha(t_L)},
% \label{eq1_1a}
% \end{equation}
\begin{equation}
\frac{v_F(x,t_F)}{\alpha(t_F)}=
\frac{\frac{v_L(x,t_L)}{\alpha(t_L)} - \frac{v_T(t_L)}{\alpha(t_L)}}{1+I'(x)[\frac{v_L(x,t_L)}{\alpha(t_L)} - \frac{v_T(t_L)}{\alpha(t_L)}]} + \frac{v_T(t_F)}{\alpha(t_F)},
\label{eq1_1b}
\end{equation}
where $t_{L(F)}$ is the physical moment at which leading (following) leaf
passes over point $x$ on the target.

{\em Optimizing irradiation dose delivered to the organ at risk:}
Ref.\cite{Papiez07:MP} investigated an algorithm based on 
a 4D DMLC leaf sequencing for the purpose 
of minimizing the irradiation dose delivered to OAR.
The optimization has been carried out by segmentation
of the velocities of leaves in case when delivery was characterized 
by constant $\alpha$. 
According to the results presented in Ref.\cite{Papiez07:MP}, it is possible
to reduce the delivered dose to OAR to 80\% of the dose minimized relative to 
the best starting point of treatment delivery during all phases of motion of 
the treatment anatomy.
Here we adopt similar strategy as presented in Ref.\cite{Papiez07:MP}. 
However with  
the inclusion of $\alpha$ as a new controllable parameter,
we illustrate that the reduction of the dose delivered 
to OAR can be improved significantly.

We formulate the problem of minimizing the irradiation dose to the
OAR by the variational principle \cite{GoldsteinBook}. 
We define the integral of irradiation dose delivered to OAR 
as an objective function (or {\em cost}-function which is required 
to be minimized)
\begin{eqnarray}
D_{OAR}[x_L, x_F, x_T, \alpha] &&\equiv \int_{\Omega_{OAR}} dx~ I_{OAR}(x)
\nonumber \\&&
= \int_{x_1}^{x_2} dx~ (m_F-m_L)(x).
\label{eq5_5}
\end{eqnarray}
Here $x$ denotes a point on OAR (in laboratory frame of reference), and
$\Omega_{OAR}\in [x_1,x_2]$ defines the effective area and boundaries of OAR.
Formally $D_{OAR}$ is a functional of $x_L,~x_F,~x_T$ and $\alpha$.
Among these parameters we can vary $x_L$ and $\alpha$ [as the 
motion of the target is not externally controllable, and
$x_F$ is assumed to be a function of $x_L$, given by Eq. \ref{eq1} or  
Eqs. (\ref{eq4}-\ref{eq5})].
Geometrically Eq. (\ref{eq5_5}) represents the area surrounded by the leaf's
trajectories and the boundaries of OAR, equivalent to the area of OAR under
radiation.
The optimal solution in form of system of Euler-Lagrange equations 
\cite{GoldsteinBook} can be determined via 
\begin{eqnarray}
\delta D_{OAR} = 0.
\label{eq5_6}
\end{eqnarray}
The functional variation has been carried out 
with respect to independent variables $x_L$, and $\alpha$.
Minimizing $D_{OAR}$ requires longer time of delivery with
respect to the treatment plans in which minimization of dose delivered to
OAR is not taken into account. 
Because analytical solutions of this problem are not achievable,
a computational method based on genetic algorithm 
is employed \cite{KozaBook}. 
% This calculation is based on numerical solutions of Eqs. (\ref{eq4}), 
% and (\ref{eq5}) in conjunction with Eqs. (\ref{eq5_5}) and (\ref{eq5_6}).

{\em Computational aspects:}
The numerical optimization is carried out in two steps.
In first step we start with an initial constant $\alpha$, and calculate  
the optimal leaf trajectories defined by velocities $v_L$, and $v_F$, 
using Eqs.(\ref{eq4}), and (\ref{eq5}).
For the optimization of velocities we divide the target into $N_{\rm seg}$
segments.
In each segment we assign a random number to $v_L$ and calculate $v_F$. 
We use genetic algorithm that 
iteratively transforms 
a population of $(v_L,v_F)$ with fixed-length $N_{\rm seg}$  
each with an associated fitness value defined as $D^{-1}_{OAR}$, 
into a new population of offspring $(v_L,v_F)$ using the Darwinian principle 
of natural selection and using operations that are patterned after 
naturally occurring genetic operations, such as mating crossover,
suppression and mutation \cite{KozaBook}.
As a result, optimal velocities corresponding to constant 
$\alpha$ are identified.

In second step we use leaf trajectories obtained in the first step,
together with the constant beam dose rate as assumed initially, 
as a starting strategy denoted by $(\alpha,v_L,v_F)$. 
In this step we calculate the optimal variable dose rate with appropriate 
leaf trajectories. 
The result of this calculation lead to an updated strategy plan  
$(\tilde{\alpha},\tilde{v}_L,\tilde{v}_F)$.
The connection between strategies in steps one and two  
can be formulated by an appropriate transformation which keeps Eq. \ref{eq1_1b}
invariant, and preserves the integrity of the dose delivered to target. 
In the target frame of reference this transformation 
is given by $\tilde{m}_L=m_L$, and $\tilde{m}_F=m_F$.
Using MU to physical time transformation, we find an equivalent transformation
for the velocities of the leaves 
\begin{eqnarray}
\frac{1}{\tilde{\alpha}}\left[\tilde{v}_L(\tilde{t}_L)-\tilde{v}_T(\tilde{t}_L)\right]
= 
\frac{1}{\alpha}\left[v_L(t_L)-v_T(t_L)\right].
\label{eq6}
\end{eqnarray}
Here $v_L(t_L)=dx_L(t_L)/dt_L$, and 
$\tilde{v}_L(\tilde{t}_L)=d\tilde{x}_L(\tilde{t}_L)/d\tilde{t}_L$.
From Eq.(\ref{eq6}) one can deduce 
\begin{eqnarray}
\tilde{x}_L(\tilde{t}_L)-\tilde{x}_T(\tilde{t}_L) = x_L(t_L)-x_T(t_L). 
\label{eq7}
\end{eqnarray}
Similar equations holds for $v_F$, and $x_F$.
Similar to the first step,  
at point $x$ on the target, delivered dose to the target is calculated by
$I(x)=\tilde{m}_F - \tilde{m}_L$.  
Here $\tilde{m}_{L(F)}=\int^{\tilde{t}_{L(F)}}_0 dt' \tilde{\alpha}(t')$ where 
the leading (following) leaf arrives at $x$ on the target at 
$\tilde{t}_{L(F)}$.
Note that the trajectories calculated in second step are required to
preserve the integrity of $I(x)$.
These equations imply that the trajectories of the leaves in target 
frame of reference, expressed in MU, are invariant under variation of 
the intensity dose rate. 
This in turn means that shape of $I(x)$ (that is 
determined by the mutual relationship between leaf velocities)
is kept invariant (integrity of delivery is preserved). 
The procedure described above, base on steps 1 and 2, can be repeated 
in a loop that continuous till it converges to a satisfactory solution.
%\cite{remark}.

\begin{figure}
\begin{center}\vspace{1cm}
\includegraphics[width=0.9\linewidth]{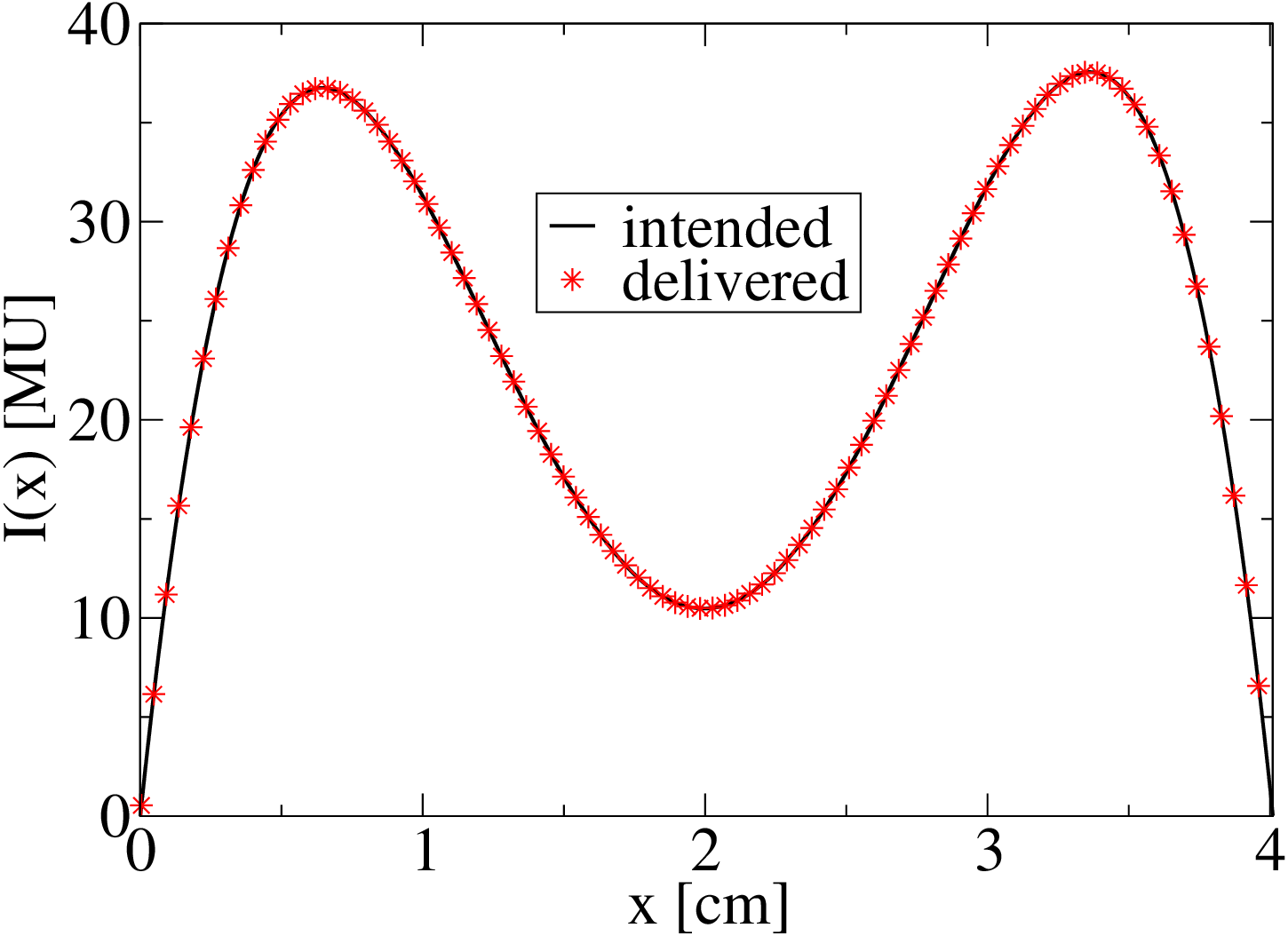}\vspace{1cm}
%~/LechPapiez/AAPM_2008/numerics/Opt_Int_rate
\caption{
The double parabola intensity used in the calculations.
The solid line indicates the intensity map intended to be delivered
and the line with stars shows the intensity 
actually delivered to the moving target.
}
\label{fig0}
\end{center}
\end{figure}

\begin{figure}
\begin{center}\vspace{1cm}
\includegraphics[width=0.9\linewidth]{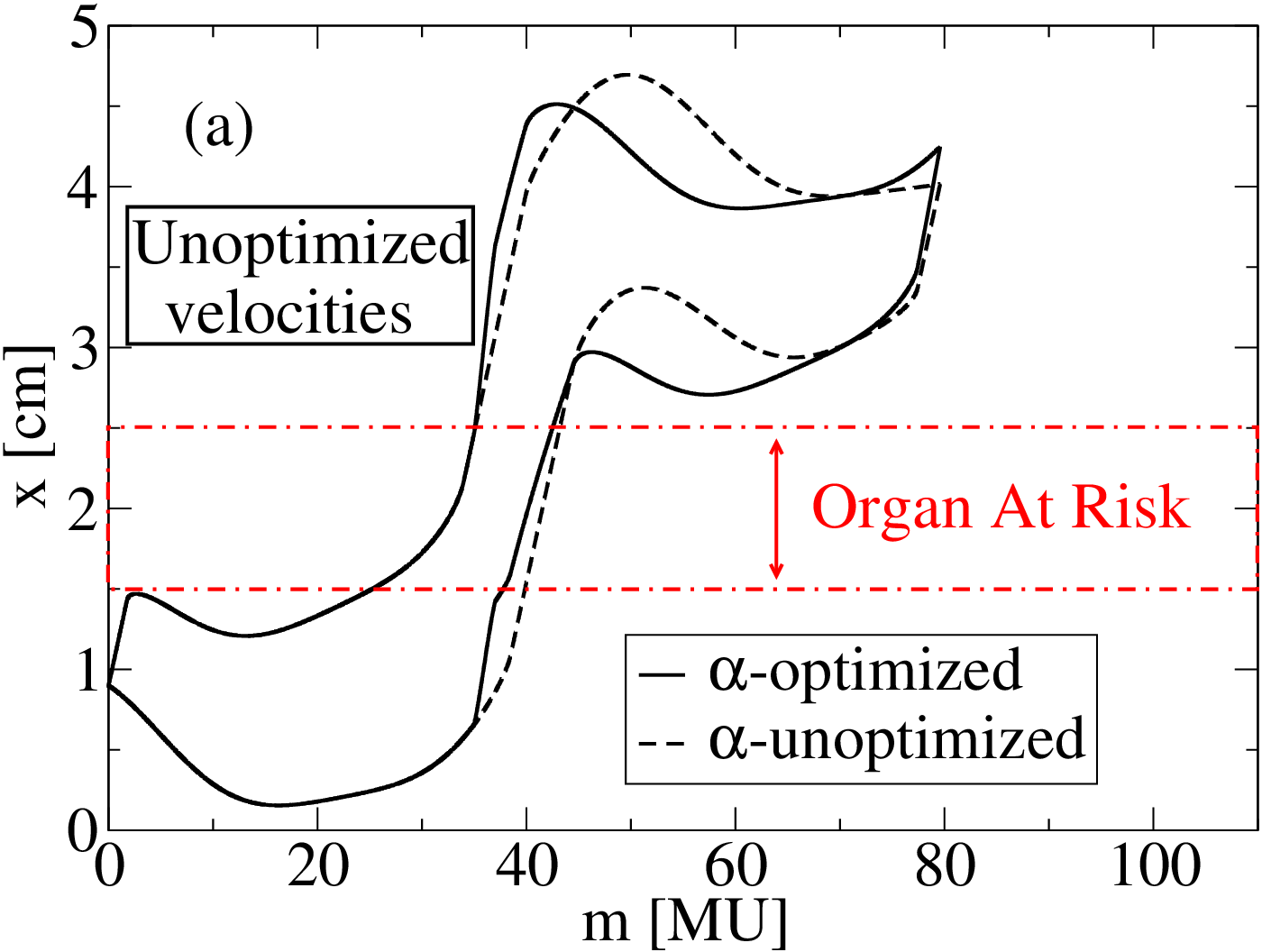}\vspace{1cm}
\includegraphics[width=0.9\linewidth]{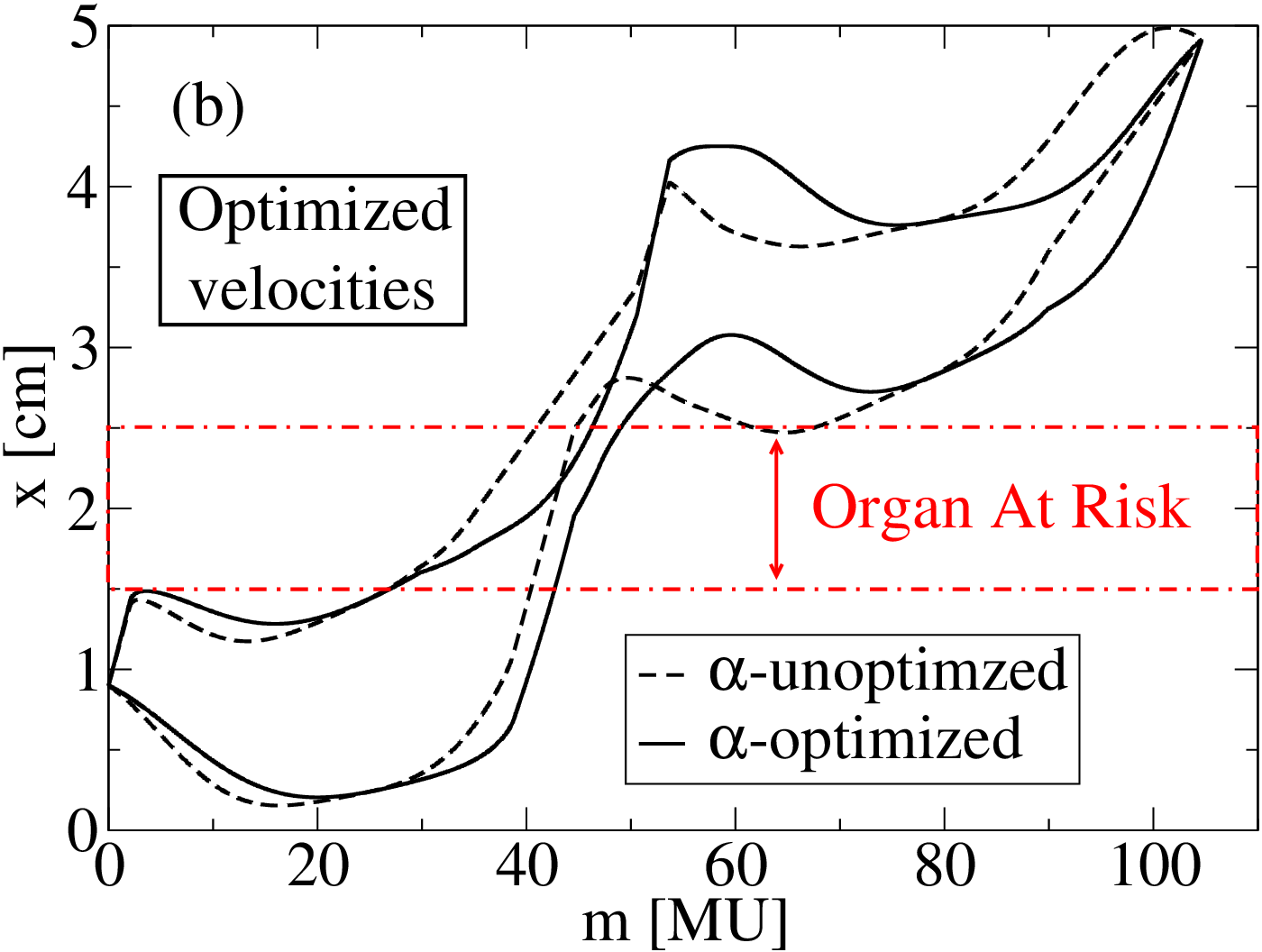}
%~/LechPapiez/AAPM_2008/numerics/Opt_Int_rate
\caption{
Trajectories of the leaves shown in the laboratory frame of reference
expressed as a functions of cumulative number of monitor units $m$.
The delivery strategy is based on optimized time of delivery and 
irradiation dose delivered to the organ at risk.
Figure (a) shows trajectories of leading and following leaves calculated 
based on uniform intensity dose rate $\alpha=10$ MU/sec (dashed line), and
$\alpha=2$ MU/sec for $3.5 \leq t \leq 4.5$ sec and $\alpha=10$ MU/sec 
elsewhere (solid line). 
The latter has been calculated based on optimization of step-wise dose rate 
that minimizes intensity to static organ at risk.
Both trajectories are calculated based on unoptimized leaf velocities. 
We observe that modulation of the intensity dose rate lowers the
irradiation dose delivered to ORA.
In figure (b) optimization of the leaf velocities has been performed by 
segmentation of the target into five pieces. 
In each piece the optimal $v_L$, and $v_F$ have been calculated using 
genetic algorithm, with $\alpha=10$ (dashed line), and optimized 
intensity dose rate (solid line).
In the latter, the self-consistent optimum intensity dose rate 
is $\alpha=13$ MU/sec for $0 \leq t \leq 2.5$ sec 
and $\alpha=10$ MU/sec for $t > 2.5$ sec. 
A comparison with a strategy based on uniform intensity dose rate 
$\alpha=10$ MU/sec with unoptimized leaf velocities 
shown in figure (a) by dashed line, 
and the self-consistent optimized leaf velocities and 
intensity dose rate shown in figure (b) by solid line, shows that
the dose delivered to the organ at risk (the area surrounded by the leaves
trajectories and the boundaries of the OAR) 
can be reduced significantly.
}
\label{fig1}
\end{center}
\end{figure}

\begin{figure}
\begin{center}\vspace{1cm}
\includegraphics[width=0.9\linewidth]{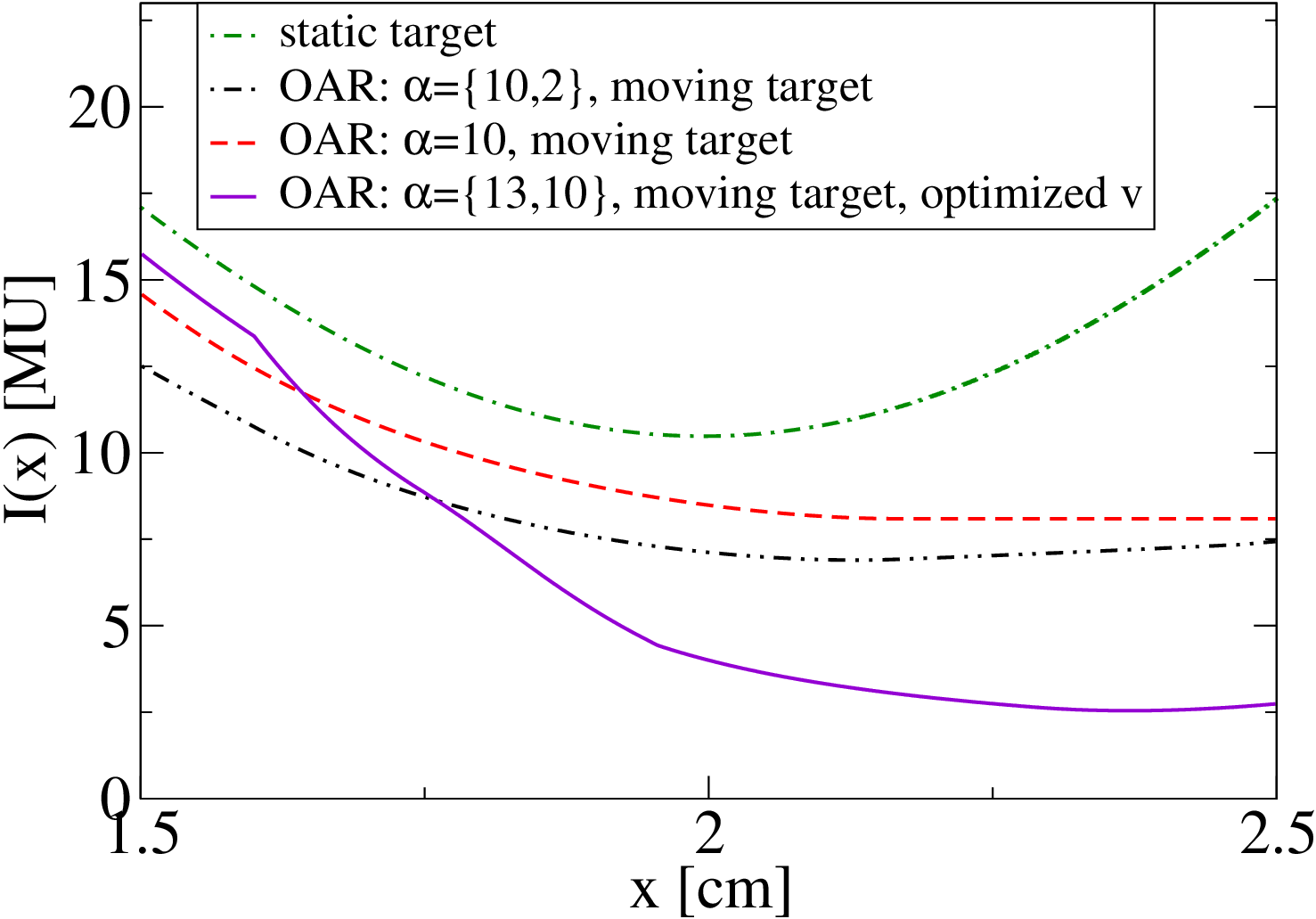}\vspace{1cm}
\caption{
Profiles of intensity delivered to 
static organ at risk 
calculated for static target (green dashed line),
optimized with respect to beam dose rate  
(black and red dashed lines), and 
optimized with respect to both of leaf velocities and beam dose rate delivery
(purple solid line).
The integral of dose delivered to the organ at risk is calculated 
for each case and is given by 12.8, 9.5, 8, and 6 MU. 
We observe that the optimization based on leaf velocities and intensity
dose rate may reduce the irradiation dose delivered to organ at risk down
to 45\%. 
}
\label{fig2}
\end{center}
\end{figure}

{\em Numerical results:}
For illustration of the method 
we consider a moving rigid target with length $x_f=4$
and a static OAR in laboratory frame of reference. 
For simplicity, the trajectory of the target is assumed to be one-dimensional 
and is given by $x_T=\sin^4(0.6 t + 1.8)$.
The dimension of OAR is assumed to be $2.5 \leq x \leq 3.5$. 
The length and time are expressed in cm and second respectively. 
The planned intensity to be delivered to the target 
is a double parabola as illustrated by solid line in Fig. \ref{fig0}.
We also consider the maximum accessible velocity of the leaves and 
intensity dose rate as $v_{\rm max}=3$ cm/sec, and 
$\alpha_{\rm max}=1000$ MU/min.
For the optimization of the leaf velocities,
we divide the target up to seven segments ($N_{\rm seg}=1$ to 7).
In each segment we associate ten random velocities to the leading leaf
subjected to $v_L\leq v_{\rm max}$.
The number of possible combinations for $v_L$ and thus the size 
of search space is given by
$10^{N_{\rm seg}}$ and exceeds up to $10^7$.
For the optimization of $\alpha$, a step-wise beam dose rate with 
one step has been assumed with four parameters: 
the initial and final moments, and the intensity dose rates of the pulse.
The optimal value of these parameters are calculated by the random 
search optimization procedure. 
The optimized delivery plans shows that the reduction of
the dose delivered to OAR is possible to be as low as 45\%
for $N_{\rm seg}=3$, 5, and 6.
For illustration we present the numerical results 
in Figs. \ref{fig1} and \ref{fig2}, calculated based on
the optimization of $v_L$ with $N_{\rm seg}=5$ corresponding to
optimal $\alpha$ given by   
\begin{eqnarray}
\alpha =
\left\{\begin{array}{c}
13~{\rm MU/sec}~~~~~~~~~~~~~~~~ 0 \leq t \leq 2.5\\
10~{\rm MU/sec}~~~~~~~~~~~~~~~~~~~~~ t > 2.5
\end{array}\right.
\label{eq8}
\end{eqnarray}

{\em Conclusion}:
% Radiotherapy is a common modality for the treatment of cancer. 
% The desirable outcome of the radiation therapy is tumor ablation 
% while avoiding damage to healthy tissues. 
% The latter is one of the most challenging problems in IMRT.
% In the conventional planning system in IMRT the body anatomy 
% is assumed to be static.
% However,
In this work we showed that the internal motion of body anatomy during therapy 
can be used as a valuable information in protecting the sensitive organs.
To minimize the error in IMRT delivery to moving targets we developed
strategies based on controlling the velocity of multi-leafs and 
intensity rate of beams that allow the irradiation of the target to 
be properly re-distributed over subsequent phases of target motion. 
This general approach utilizes multiple solutions that provide the 
same dose to target while optimizing these delivery parameters that 
can improve the quality of the therapy. 
% The solution strategies involve complex optimization processes on 
% diverse landscapes with multiple local minima. 
% We explicitly explore various dynamical pathways and use the genetic 
% algorithm to arrive at efficient solutions of our multi-parameter 
% optimization problems. 
% In conclusion we showed that by controlling 
% the leaf velocities and intensity dose rates
% it is possible to protect the organs at risk
% from the excessive irradiation in IMRT treatment 
% planning.
Our calculation based on optimization of the control parameters 
using genetic algorithm
shows that the dose delivered to the organ at risk can 
be reduced significantly
compare to the delivery based on static anatomy assumption.

% \begin{eqnarray}
% \label{eq1}
% \end{eqnarray}

%%%%%%%%%%%%%%%%%%%%%%%%%%%%%%%%%%%%%%%%%%%%%%%%%%%%%%%%%%%%%%%%%%%%%%%%%%%%%%%
       
%%%%%%%%%%%%%%%%%%%%%%%%%%%%%%%%%%%%%%%%%%%%%%%%%%%%%%%%%%%%%%%%%%%%%%%%%%%%%%%

%%%%%%%%%%%%%%%%%%%%%%%%%%%%%%%%%%%%%%%%%%%%%%%%%%%%%%%%%%%%%%%%%%%%%%%%%%%%%%%

\end{document}